\documentclass[twocolumn,showpacs,showkeys,preprintnumbers,superscriptaddress,amsmath,floatfix,amssymb,secnumarabic,nofootinbib]{revtex4}
\usepackage[colorlinks=true]{hyperref}
\usepackage{graphicx}
\usepackage{epsfig}

\def\({\left(}
\def\){\right)}
\def\[{\left[}
\def\]{\right]}

\newcommand{\lr}[1]{ \left( #1 \right) }
\newcommand{\lrs}[1]{ \left[ #1 \right] }

\newcommand{\expa}[1]{ \exp{\left( #1 \right)} }

\newcommand{\beq} {\begin{eqnarray}}
\newcommand{\eeq} {\end{eqnarray}}

\newcommand{\comment}[1]{}

\newcommand{\logo}{\\ \vskip -18mm
\leftline{\includegraphics[scale=0.3,clip=false]{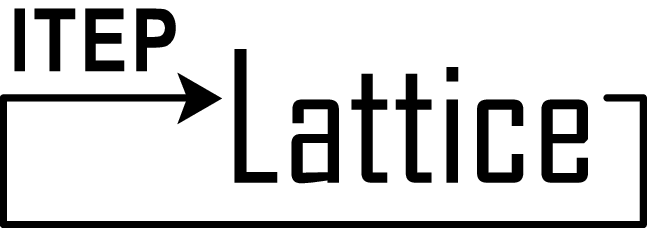}} \vskip 10mm}

\begin{document}
\sloppy

\title{Interaction of static charges in graphene within Monte-Carlo simulation \logo}

\author{V.~V.~Braguta}
\email{braguta@itep.ru}
\affiliation{Institute for High Energy Physics, Protvino, 142281 Russia}
\affiliation{Institute of Theoretical and Experimental Physics, Moscow, 117218 Russia}

\author{S.~N.~Valgushev}
\email{semuon06@gmail.com}
\affiliation{Institute of Theoretical and Experimental Physics, Moscow, 117218 Russia}
\affiliation{Moscow Inst Phys \& Technol, Institutskii per. 9, Dolgoprudny, Moscow Region, 141700
Russia}

\author{A.~A.~Nikolaev}
\email{nikolauev@gmail.com}
\affiliation{Far Eastern Federal University, Vladivostok, 690091 Russia}

\author{M.~I.~Polikarpov}
\email{polykarp@itep.ru}
\affiliation{Institute of Theoretical and Experimental Physics, Moscow, 117218 Russia}
\affiliation{Moscow Inst Phys \& Technol, Institutskii per. 9, Dolgoprudny, Moscow Region, 141700
Russia}

\author{M.~V.~Ulybyshev}
\email{ulybyshev@goa.bog.msu.ru}
\affiliation{Institute of Theoretical and Experimental Physics, Moscow, 117218 Russia}
\affiliation{Institute for Theoretical Problems of Microphysics, Moscow State University, Moscow, 119899 Russia}

\begin{abstract}

The study of the interaction potential between static charges  within Monte-Carlo
simulation of graphene is carried out. The numerical simulations are performed in the effective
lattice field theory with noncompact $3 + 1$-dimensional Abelian lattice gauge fields and
$2 + 1$-dimensional staggered lattice fermions. It is shown that for all considered  temperatures
the interaction can be well described by the Debye screened potential created by two-dimensional electron-hole
excitations. At low temperatures Debye mass $m_D$ plays a role of order parameter of the insulator-semimetal
phase transition.
In the semimetal phase at high temperature graphene reveals the properties
of weakly interacting two-dimensional plasma of fermions excitations.

\end{abstract}
\pacs{05.10.Ln, 71.30.+h, 72.80.Vp}
\keywords{graphene, electron transport, Coulomb interaction, Monte-Carlo simulations}

\maketitle

\section{Introduction}
\label{IntroductionSec}

 Graphene is an allotrope of carbon, in which atoms form a two-dimensional honeycomb lattice. Carbon atoms in it are bonded by $sp^2$-bonds and the bond length is about 0.142 nanometers \cite{Raji:1}.

 The charge carriers in graphene behave as massless fermions \cite{semenoff}. The Fermi velocity of charge carriers is $v_F\approx\frac{c}{300}$.
Since the Fermi velocity is much smaller than the speed of light, magnetic and retardation effects  in the interactions between charge carriers may be neglected, thus electron-electron interaction in graphene is well described by the instantaneous Coulomb potential.  The effective coupling constant for the Coulomb interaction in graphene
$\sim \frac{e^2}{v_F}\approx 2$ is large, so this material can be considered as a strongly interacting system.

In real experiments graphene is put on a substrate. The effective coupling constant
for graphene on substrate with the dielectric permittivity $\epsilon$
is reduced by a factor $2/(\epsilon+1)$. The variation of the dielectric
permittivity $\epsilon$ of  substrate changes effective coupling constant and thus allows
to study the properties of graphene in strong and week coupling regime.

In the weak coupling regime theoretical description of graphene properties based on perturbation
theory gives reliable results. In strong coupling regime there are no accurate analytical approaches and
Monte-Carlo simulation is an adequate method to study graphene
in strong coupling.

There exists a number of papers where graphene was studied by Monte-Carlo method
\cite{Lahde:09:1, Hands:08:1, Buividovich:2012uk, Buividovich:2012nx} and
insulator-semimetal phase transition was found.
At weak coupling regime graphene is in the semimetal phase. In this phase the conductivity is $\sigma \sim e^2/h$ and
there is no gap in the spectrum of fermionic excitations. The chiral symmetry of graphene
is not broken. At strong coupling regime graphene is in the insulator phase. In this phase the conductivity is
considerably suppressed, there is an energy gap in the spectrum of fermionic excitations, and fermionic
chiral condensate $\langle \bar \psi \psi \rangle$ is not zero.
The phase transition from weak  to strong coupling regime takes place at
the dielectric permittivity of substrate $\epsilon \sim 4$.

In this paper we study the interaction potential between static charges in graphene for various values of
the dielectric permittivity of substrate $\epsilon$ and the temperature of graphene charge carriers $T$.
\footnote{We discuss
phenomena related to electron degrees of freedom and  neglect the thermal
vibration of the graphene honeycomb lattice. Thus we can consider the temperatures $T \sim 10^3 - 10^4$ K, at which the real graphene is melted.}.
We present the results of MC simulations of graphene in the framework of effective field model. The non-MC calculations of the potential were performed in  \cite{Katsnelson} (see also references therein).

The paper is organized as follows. In the next section a brief review of the simulation algorithm is given.
In the last section the results of numerical simulations are presented and discussed. In Appendix we derive the potential of Debye screening for two-dimensional plasma.

\section{ Lattice simulation of graphene }

\subsection{Simulation algorithm.}
\label{subsec:basic_defs}

The partition function of graphene effective field theory can be written as  \cite{Novoselov:04:1, Novoselov:09:1,
Geim:07:1, semenoff}
\begin{eqnarray}
\label{partfan}
 \mathcal{Z}   =
 \int \mathcal{D}\bar{\psi} \mathcal{D}\psi \mathcal{D}A_0
 \exp\left( -\frac{1}{2}\int d^4x \lr{\partial_{i} A_{0}}^2
 - \right. \nonumber\\ \left. -
 \int d^3x \, \bar{\psi}_f \, \lr{ \Gamma_0 \, \lr{\partial_0 - i g A_0}
 - \sum\limits_{i=1,2}
    \Gamma_i \partial_i} \psi_f
 \right) ,
\end{eqnarray}
where $A_{0}$ is the zero component of the vector potential of the $3 + 1$ electromagnetic field, $\Gamma_{\mu}$  are Euclidean gamma-matrices and
$\psi_f$ ($f = 1, 2$) are two flavors of Dirac fermions which correspond to  two spin components of the non-relativistic electrons in graphene,
effective constant $g^2 = 2 e^2/( v_F (\epsilon+1))$ ( $\hbar=c=1$ is assumed ).

The zero component  of the vector potential
$A_0$ satisfies periodic boundary condition in space and time $A_{0} (t=0)=A(t=1/T)$, where $T$ is temperature. The
fermion spinors  satisfy periodic boundary condition in space and antiperiodic boundary condition in the time
direction $\psi_f(t=0)=-\psi_f(t=1/T)$.
Partition function (\ref{partfan}) doesn't depend on the vector part of the gauge potential $A_i$, since we are
working at the leading approximation in $v_F$.

The simulation of partition function (\ref{partfan}) is carried out within the approach developed in  \cite{Lahde:09:1, Buividovich:2012uk}.
In order to discretize the fermionic part of the action in (\ref{partfan})  staggered fermions \cite{MontvayMuenster, DeGrandDeTarLQCD} are used.
One flavor of staggered fermions in $2 + 1$   dimensions corresponds to two flavors of continuum Dirac fermions \cite{ MontvayMuenster, DeGrandDeTarLQCD, Burden:87:1},
which makes them especially suitable for simulations of the graphene effective field theory.

 The action for staggered fermions coupled to Abelian lattice gauge field is

\begin{eqnarray}
\label{lat_act_interact}
S_{\Psi}\lrs{\bar{\Psi}_x, \Psi_x, \theta_{x, \, \mu}} =
 \sum\limits_{x, y} \bar{\Psi}_x \, D_{x, y}\lrs{\theta_{x, \, \mu}} \,
\Psi_y
= \nonumber \\ =
  \, \sum\limits_{x} \, \delta_{x_3, \, 0} \, \left(
\sum\limits_{\mu=0, 1, 2}
 \frac{K_\mu}{2} \bar{\Psi}_x \alpha_{x, \mu} e^{i \theta_{x, \, \mu}} \Psi_{x+\hat{\mu}}
 - \right. \nonumber \\ \left. -
 \sum_{\mu=0, 1, 2} \frac{K_\mu}{2}
 \bar{\Psi}_x \alpha_{x, \mu} e^{-i \theta_{x, \, \mu}} \Psi_{x-\hat{\mu}}
 + m {\bar{\Psi}}_x \Psi_x \right) ,
\end{eqnarray}
where $K_\mu=1$ for links in spatial directions ($\mu=1,2$) and
$K_\mu={a_s}/{a_t}$ for links in time direction ($\mu=0$),
$a_s$ and $a_t$ are the spatial and temporal lattice spacings,
the lattice coordinates $x^{\mu} = 0 \ldots L_{\mu}-1$ ( $L_1=L_2=L_3=L_s$ ), and $x^3$ is restricted to $x^3 = 0$ in the fermionic action,
 $\bar{\Psi}_x$ is a single-component Grassman-valued field, $\alpha_{x, \mu} = (-1)^{x_0 + \ldots + x_{\mu-1}}$, and $\theta_{x, \, \mu}$ are the link variables which are the lattice counterpart of the vector potential $A_{\mu}\lr{x}$.

It should be noted that nonzero mass term in (\ref{lat_act_interact}) is necessary in order
to ensure the invertibility of the staggered Dirac operator $D_{x, y}$. Physical results
are obtained by extrapolation of the expectation values of physical observables to the limit $m \rightarrow 0$.

To discretize the electromagnetic part of partition function (\ref{partfan}) the noncompact action is used
\begin{eqnarray}
\label{gauge_lat_act}
 S_g\lrs{\theta_{x, \, \mu}} = \frac{\beta}{2} \, \sum\limits_x \sum\limits^{3}_{i=1}
 \lr{ \theta_{x, \, 0} - \theta_{x + \hat{i}, \, 0} }^2  ,
\end{eqnarray}
where the summation is carried out over all 4D lattice. The constant $\beta$ is defined as follows
\begin{eqnarray}
\label{lattice_coupling_constant}
 \beta = \frac{v_F}{4 \pi e^2} \, \frac{\epsilon + 1}{2} \biggl ( \frac {a_s} { a_t} \biggr ).
\end{eqnarray}
The factor $\frac{\epsilon + 1}{2}$ takes into account the electrostatic screening for graphene on the
 substrate.

 Since action (\ref{lat_act_interact}) is bilinear in fermionic fields, they can be integrated out
\begin{eqnarray}
\label{ferm_integrated}
 \mathcal{Z} =
 \int \mathcal{D}\bar{\Psi}_x \, \mathcal{D}\Psi_x \, \mathcal{D}\theta_{x, \, 0}\,
 \nonumber \\
 \exp\lr{ - S_g\lrs{\theta_{x, \, 0}} - S_{\Psi}\lrs{\bar{\Psi}_x, \Psi_x, \theta_{x, \, 0}}}
 = \nonumber \\ =
 \int \mathcal{D}\theta_{x, \, 0}\,
  \expa{ -S_{eff}\lrs{\theta_{x, \, 0}} }  ,
\end{eqnarray}
where
\begin{eqnarray}
\label{eff_action}
 S_{eff}\lrs{ \theta_{x, \, 0} } = S_g\lrs{ \theta_{x, \, 0} } - \ln \det\lr{ D\lrs{ \theta_{x, \, 0} } } .
\end{eqnarray}

To generate the gauge field configurations
with the statistical weight $\expa{-S_{eff}\lrs{ \theta_{x, \, 0} }}$
the standard Hybrid Monte-Carlo Method is used \cite{MontvayMuenster, DeGrandDeTarLQCD, Lahde:09:1}.
 In order to speed up the simulations we also perform local heatbath updates of the gauge field outside of the graphene plane (at $x^3 \neq 0$) between Hybrid Monte-Carlo updates.
 Both algorithms satisfy the detailed balance condition for the weight (\ref{ferm_integrated}) \cite{MontvayMuenster, DeGrandDeTarLQCD} and the path integral weight (\ref{ferm_integrated}) is the stationary probability distribution for such a combination of both algorithms. Since heatbath updates are computationally very cheap, they significantly decrease the autocorrelation time of the algorithm.

The temporal lattice spacing $a_t$ is equal to the spatial lattice spacing $a_s$ in symmetric lattice.
As was explained before to take into account the Coulomb interaction between quasiparticles in graphene
it is sufficient to introduce only the fourth component of electromagnetic vector potential. This might
imply that discretization in temporal direction is particularly important to get reliable results.
In the calculation we fix the temperature of graphene sample and vary the discretization in the temporal
direction in order to address this point in detail.

In the simulation of effective theory (\ref{partfan}) the lattice spacing $a_s$ plays a role of ultraviolet cut off.
The exact value of this cut off is unknown. One can only state that $a_s \sim 0.142$ nanometers,
which is the distance between two neighbouring carbon atoms in graphene.
To clarify the physical scale of dimensional quantities we put their
values assuming that $a_s=0.142$ nanometers. In addition, in brackets we put
 dimensional parameters in terms of $\sim 1/a_s$.

\subsection{Physical observables on the lattice}
\label{subsec:observables}

\begin{figure}[t]
 \includegraphics[width =5.5cm, angle =270]{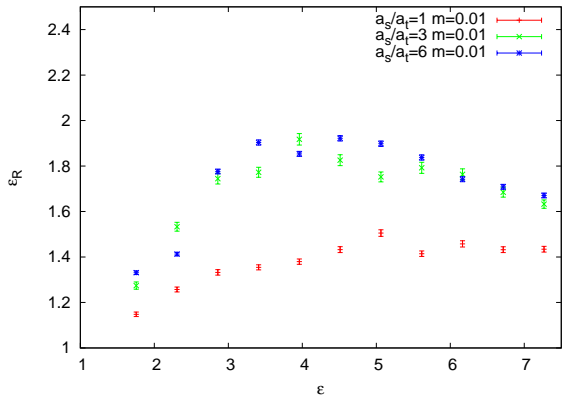}\\
 \caption{The dielectric permittivity of graphene $\epsilon_R$
as a function of the dielectric permittivity of substrate $\epsilon$ for different $a_s/ a_t$ at $ T=0.23 eV (a_s T=0.00019)$. }
 \label{fig:discr}
\end{figure}

  To measure the potential, $V(r)$, between static charges, we calculate the correlator
of two Polyakov lines $\langle P^{\gamma}(0) (P^{\gamma}(\vec r))^+ \rangle$:
\beq
\langle P^{\gamma}(0) (P^{\gamma}(\vec r))^+ \rangle = a \exp { \biggl ( - \frac {V( \vec r)} T \biggr ) }.
\eeq
where $T$ is the temperature of graphene sample, the Polyakov line $P(\vec r)$ is
\beq
P(\vec r) = \exp { \bigl ( - i e \int_0^{1/T} d t A_0 (t, \vec r)  \bigr )  } = \prod_{t=0}^{L_t-1} \exp {  ( - i  \theta_{(t,\vec r), 0}  )  }
\eeq

To suppress statistical errors, we measure the correlator of Polyakov lines in some rational power.
Physically this means that the interaction potential between static charges $\pm e \cdot \gamma$ is considered.
We have found that for $\gamma \sim 0.1$ the uncertainty of the calculation is much smaller than that
in the case of $\gamma=1$ (usual Polyakov line). Below the value $\gamma=0.1$ is used.


\begin{figure}[t]
 \includegraphics[width =5.5cm, angle =270]{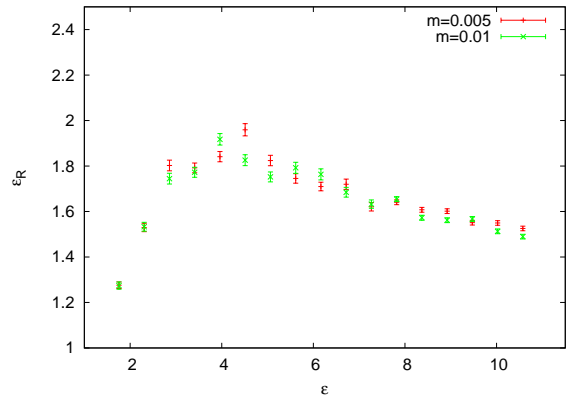}\\
 \caption{The dielectric permittivity of graphene $\epsilon_R$
as a function of the $\epsilon$ for the fermion masses $m=0.005,~ 0.01$ at $T=0.23 eV (a_s T=0.00019)$. }
 \label{fig:mass}
\end{figure}

 Below we use the following notations:
\beq
\alpha_0 = e^2  \frac 2 {\epsilon+1}
\eeq
is the bare effective charge and
\beq
\alpha_R = \frac {\alpha_0} {\epsilon_R}
\eeq
is the effective charge, renormalized due to interaction,
$\epsilon_R$ is effective dielectric permittivity of graphene.

\section{Numerical results and discussion}
\label{sec:num_results}

\subsection{The interaction potential at low temperatures}

To get the potential between static charges, we measure the correlator of Polaykov lines
 and fit $V(r)$ by lattice screened Coulomb potential:
\beq
V(\vec r) &=& \frac 1 {\epsilon_R} V_C(\vec r) + c,   \\ \label{Coul_pot}
V_C(\vec r) &=&- \alpha_R \frac {\pi \gamma^2} {L_s^3 a_s} \sum_{n_1,n_2,n_3} \frac 1 { \sum_i \sin^2 (p_i a_s/2)} e^{i \vec p \vec r},~\\
\nonumber p_i &=& \frac {2 \pi} {L_s a_s} n_i.
\nonumber
\eeq
and determine $\epsilon_R$. In formula (\ref{Coul_pot})
 $c$ is the constant, which parameterizes selfenergy contribution to the potential, $V_C(\vec r)$ is the lattice Couloumb
potential, which takes into account spatial discretization
and finite volume effects, $n_i$ are integers which run in the interval $(0,L_s-1)$ and
point $n_1=n_2=n_3=0$ is excluded.


Firstly we discuss the systematic errors due to the temporal discretization. Using the algorithm described
above we generated $100$ statistically independent gauge field configurations at the lattices
$20^3 \times L_t, L_t=20, 60, 120$ for a set of values of the dielectric permittivity of substrate $\epsilon \in (1, 8)$.
These three lattices correspond to the temperature $T=0.23 eV (a_sT=0.00019)$ and the ratios $a_s/a_t = 1, 3, 6$ correspondingly.
We have found an excellent agreement between our data and expression  (\ref{Coul_pot}) ($\chi^2/dof \sim 1$ for all $\epsilon$).
Thus this result confirms that {\it static charges at low temperature in graphene interact via Coulomb potential.}

The dielectric permittivity of graphene $\epsilon_R$
as a function of the dielectric permittivity of substrate $\epsilon$ for different $a_s/a_t$ is
shown in Fig. \ref{fig:discr}. From this plot one sees that there is large difference between the
results obtained at $a_s/a_t=1$ and $a_s/a_t=3$. At the same time the results for
$a_s/a_t=3$ and $a_s/a_t=6$ are in a reasonable agreement with each other. It seems that at $a_s/a_t \sim 3-6$
one approaches to the continuum limit in the temporal direction. Below $a_s/a_t=6$ discretization scheme is used.

Now let us turn to the fermion mass dependence of our results. In Fig. \ref{fig:mass}  $\epsilon_R$  as a function of the $\epsilon$
for the fermion masses $m=0.005, 0.01$ is shown. Within the uncertainty of the calculation
the results obtained for different masses are compatible to each other. The simulation of the
gauge field configurations with the mass $m=0.005$ is much more time consuming as compared to the
mass $m=0.01$. So, to decrease the time of the calculation, all calculations are carried out for
the fermion mass $m=0.01$. In addition to the fermion mass dependence, we studied the volume dependence of our results and
found that this dependence is very weak.

\begin{figure}[t]
 \includegraphics[width =5.5cm, angle =270]{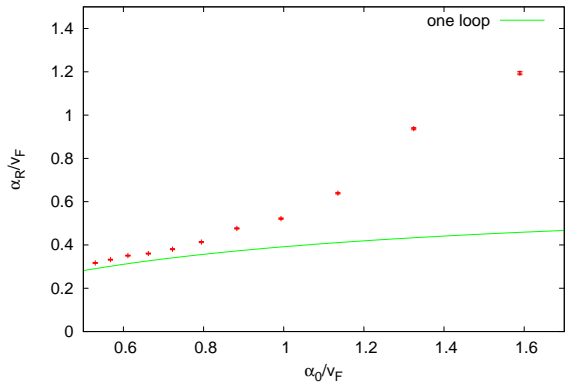}\\
 \caption{ The renormalized charge squared $\alpha_R$ as a function of the bare charge squared
$\alpha_0$ rescaled by the $v_F$ and the plot of one loop formula (\ref{one_loop}).
The insulator-semimetal phase transition takes place at $\alpha_0/v_F \sim 0.9$. }
 \label{fig:renorm}
\end{figure}

In Fig. \ref{fig:renorm}
we show how $\alpha_R$ is renormalized due to the interaction.
In the semimetal
phase the effective coupling constant is not large $\alpha_0/v_F < 1$ and one can try to apply perturbation theory
to disribe our data.  At one loop approximation the dependence of $\alpha_R$ on the $\alpha_0$ for graphene  is given by the expression \cite{Gonzalez:1993uz}
\beq
\frac {\alpha_R} {\alpha_0} = \frac 1 {1 + \frac {\pi} 2 \frac {\alpha_0} {v_F} } = \frac 1 {1 + 3.4  \frac 2 {\epsilon+1}}.
\label{one_loop}
\eeq
Fig. \ref{fig:renorm} shows that at small $\alpha_0$ we have good agreement with perturbation theory.

\subsection{The temperature dependence of the interaction potential}

To study the dependence of the dielectric permittivity $\epsilon_R$ on the temperature,
we generated $100$ statistically independent gauge field configurations at the lattices
$20^3 \times L_t, L_t=$56, 50, 38, 28, 26, 22, 18. These lattices correspond to the temperatures
$T=0.50$ eV ($a_sT=0.00041$), $T=0.56$ eV ($a_sT=0.00046$), $T=0.74$ eV ($a_sT=0.00061$), $T=1.00$ eV ($a_sT=0.00082$),
$T=1.08$ eV ($a_sT=0.00089$), $T=1.28$ eV ($a_sT=0.00105$), $T=1.56$ eV ($a_sT=0.00128$) correspondingly.

\begin{figure}[t]
 \includegraphics[width =5.5cm, angle =270]{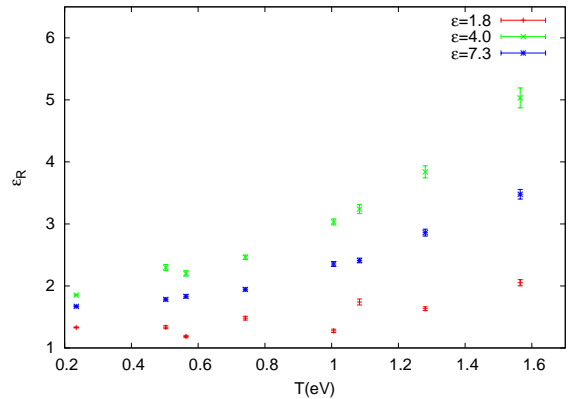}\\
 \caption{ The dependence of the $\epsilon_R$ on the temperature of graphene sample
for the $\epsilon=1.8$ (insulator phase),  $\epsilon=4.0$ (transion region), $\epsilon=7.3$
(semiconductor phase), is shown }
 \label{fig:temp}
\end{figure}

\begin{figure}[t]
 \includegraphics[width =5.5cm, angle =270]{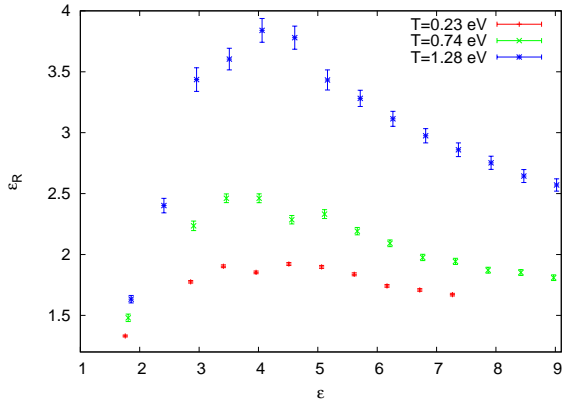}\\
 \caption{ The dependence of the $\epsilon_R$ on the $\epsilon$ at different temperatures obtained from the fitting
 with Coulomb potential $V_C(\vec r)$ (\ref{Coul_pot}).}
 \label{fig:dif_temp}
\end{figure}

In Fig. \ref{fig:temp} the dependence of the $\epsilon_R$ on the temperature of graphene sample for
different dielectric permittivities of substrate $\epsilon$ is shown. Graphene with $\epsilon=1.8$
is  in the insulator phase. It is seen that the temperature dependence in this phase is the weakest
as compared to the $\epsilon=4.0$ and $\epsilon=7.3$ points. This happens since in the
insulator phase fermions have dynamically generated mass. If this mass is larger than the temperature,
the production of free charges which enhances the $\epsilon_R$ is suppressed. If the fermions are
massless, free charges production is no longer suppressed and the temperature dependence of the $\epsilon_R$
is stronger. This effect is seen  in the semimetal phase at $\epsilon=7.3$, where
quasiparticles are massless. The most rapid temperature dependence takes place for $\epsilon=4.0$ which is in the
transition region. In this region graphene is in the insulator phase at low temperature and
in the semimetal phase at high temperature what explains the most rapid temperature dependence.
In  Fig. \ref{fig:dif_temp} the dependence of $\epsilon_R$ on $\epsilon$ at different temperatures is shown.

Formula (\ref{Coul_pot}) fits data satisfactory ( $\chi^2/dof \sim 1-3$ ) for all temperatures.
However, the larger the temperature the larger $\chi^2/dof$.
One can assume that the worsening of the fitting model can be assigned to the following fact. At
sufficiently large temperature graphene contains equal number of electrons and holes. If one puts
electric charge to such media, a nonzero charge density is created. This charge density leads
to some sort of Debye screening in graphene which is not accounted in (\ref{Coul_pot}).

In Appendix \ref{appendix} the derivation of the Debye screening in graphene  is given.
It is assumed that the interaction between quasiparticles is weak, which is the case
only for sufficienty large $\epsilon$. However, the Debye potential  (\ref{deb_cont})
without explicit expression for Debye screening mass $m_D$ (\ref{deb_mass})
can be thought of as a modification of the Coulomb potential with unknown parameter $m_D$.
In this sence formula (\ref{deb_cont}) can be applied for all values of $\epsilon$ and temperature.

\begin{figure}[t]
 \includegraphics[width =5.5cm, angle =270]{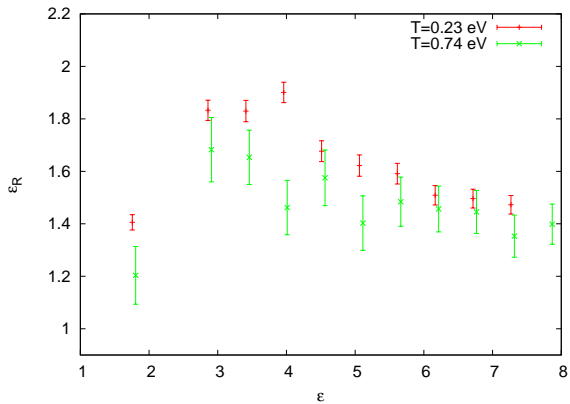}\\
 \caption{ The dependence of the $\epsilon_R$ on the $\epsilon$ at different temperatures obtained from the fitting
 with Debye screening potential $V_D(\vec r)$ (\ref{deb_lat}).}
 \label{fig:dif_temp1}
\end{figure}

To carry out the study of the temperature dependence of the interaction potential we replace
the lattice Coulomb potential $V_C(\vec r)$  by the lattice version of Debye screening potential (\ref{deb_lat}) in model (\ref{Coul_pot}).
Before the modifications of the potential the description ($\chi^2/dof > 1-3$) of the
available data was not as good  as it became
after the modification ( $\chi^2/dof < 1$ ) for all temperatures and $\epsilon$.
In Fig. \ref{fig:dif_temp1}  we plot the $\epsilon_R$ as a function of the $\epsilon$ for different temperatures.
It is seen from this plot that
contrary to the fitting procedure with Coulomb potential the $\epsilon_R$ with Debye screening
potential is almost temperature independent. So, the fitting with Debye screening potential cancels the temperature dependence from the
dielectric permittivity $\epsilon_R$ and encodes it into Debye mass $m_D$. This confirms
that in some region the temperature dependence of the interaction potential results
from the Debye screening.

\begin{figure}[t]
 \includegraphics[width =5.5cm, angle =270]{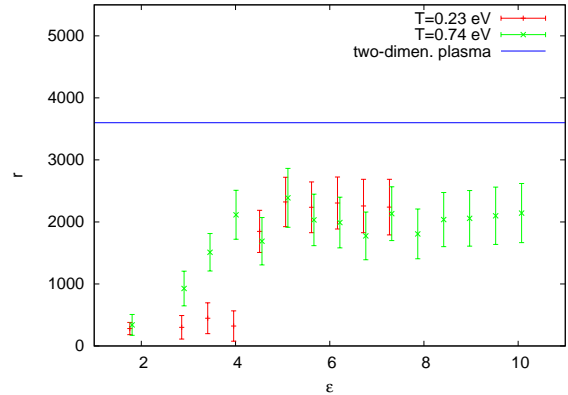}\\
 \caption{ The dependence of the ratio $r=(m_D e^2)/(T \alpha_R)$ on the $\epsilon$ at different temperatures.
The line parallel to the $\epsilon$-axis is the value of the ratio $(m_D e^2)/(T \alpha_R)$ at the approximation
of weakly interacting  two-dimensional plasma of quasiparticles.}
 \label{fig:dif_temp2}
\end{figure}

Now let us turn to Debye screening mass. Equation (\ref{deb_mass1}) defines the Debye mass for
two-dimensional plasma of quasiparticles when interaction between quasiparticles
is disregarded. It not difficult do derive the expression for Debye mass $m_D$
which is valid for the interacting quasiparticles. Evidently, if there is no
interaction between quasiparticles, $m_D=0$. This means that the expansion of $m_D$
starts for the term proportional to the $\sim \alpha_R$, which determines the
strength of the interaction. The second property
of the $m_D$ is that it disappears if the density of quasiparticles $n$ is zero. So, one concludes that
$m_D \sim n/T$, where  temperature appeared in the denominator for the dimensional reasons\footnote{The density
$n$ in graphene has dimension $\sim$(energy)$^2$.}.
Now we have the following expression
\beq
m_D = k(e^2, T)~  {\alpha_R} \frac n T,
\eeq
where $k(e^2, T)$ is some function which could depend on the $e^2$ and $T$.
In Fig. \ref{fig:dif_temp2} we present  the following observable
$r=(m_D e^2)/(T \alpha_R)$  which is proportional to the $n/T^2$. This observable allows to
study the density of quasiparticles in graphene. If the interaction between
quasiparticles is weak, the ratio $(m_D e^2)/(T \alpha_R)$  equals to
\beq
r  = \frac {m_D e^2} {T \alpha_R} = 8 \log 2 \frac {e^2} {v_F^2} \simeq 3600.
\label{const}
\eeq
In Fig.\ref{fig:dif_temp2} the dependence of the ratio $(m_D e^2)/(T \alpha_R)$ on the $\epsilon$ at
different temperatures is shown. The line parallel to the $\epsilon$-axis is the value
of the ratio $(m_D e^2)/(T \alpha_R)$  (\ref{const}).

Now few comments are in order

\begin{itemize}
\item First let us consider the semimetal phase $\epsilon>5$. In this region the ratio
$(m_D e^2)/(T \alpha_R)$ tends to some constant value and
this value is by a factor $\sim 1.5-2.0$ smaller than that given by formula (\ref{const}).
The possible source of this disagreement is that in formula (\ref{const}) we used the bare
 Fermi velocity $v_F$. Evidently one should use the renormalized Fermi-velocity $v_F^R$,
which beyond the scope of this paper. The $v_F^R$ is larger than the $v_F$, so
the inclusion of  Fermi velocity renormalization will push the constant (\ref{const}) to the
correct direction. Accounting this fact one can conclude that within the uncertainty of the
calculation {\it in the semimetal phase
electron excitations in graphene form a weakly interacting two-dimensional plasma.}

\item Assuming that the difference between constant (\ref{const}) and the position
of the plateau in Fig. \ref{fig:dif_temp2} results from Fermi velocity
renormalization one can estimate the ratio $v_F^R/v_F$ in the semimetal phase as $\sim 1.2-1.4$. This value is
in a reasonable agreement with the results obtained within Monte-Carlo
simulation of graphene \cite{Drut:2013wqa} and with experiment
\cite{exp}.

\item It is seen from Fig. \ref{fig:dif_temp2} that {\it at low temperature Debye mass $m_D$
plays a role of order parameter of the insulator-semimetal phase transition.}
At small dielectric permittivity of substrate,  $m_D$  equals zero within
the accuracy of the calculation, what means that the interaction potential
is Coulomb. At $\epsilon \sim 4-5$ Debye mass becomes nonzero,
abruptly reaching the regime of two-dimesional plasma.
The interaction in this region is due to Debye potential.  Thus the study of Debye screening mass
allows to determine the position of the insulator-semimetal phase transition, which
takes place at $\epsilon \sim 4-5$, in accordance with the results
of papers \cite{Lahde:09:1, Buividovich:2012uk}. At large temperatures $m_D$ is not zero for any values of the $\epsilon$.
It is smoothly rising function of $\epsilon$ which is saturated at $\epsilon \sim 4-5$.

\item To understand the behaviour of the Debye mass, which is proportional to the density of excitations $n$,
one can use the following model.
 In the insulator phase $\epsilon<4$ the fermion excitation acquire
dynamical mass. So, the density of charged fermion excitations is exponentially suppressed
\beq
\frac {n} {T^2} \sim \exp { \biggl ( -\frac {M_f(g^2)} {T} \biggr )}.
\eeq
The dynamical fermion mass $M_f(g^2)$ depends on the effective coupling constant $g^2=\alpha_0/v_F$.
It is seen from Fig. \ref{fig:dif_temp2}  that at temperature $T=0.23 eV (a_sT=0.00019)$  the density is either
considerably suppressed or equal to zero, what implies that $M_f(g^2)>T$. However, at temperature
$T=0.74 eV (a_sT=0.00061)$ the density is no longer suppressed and it is monotonically rising function of the
effective constant, what implies that $M_f(g^2)<T$. So, the dynamically generated  fermion mass in the insulator
region can be estimated as $M_f(g^2) \sim 0.5 eV$.

\end{itemize}

At the end of this section it worth to note that because of the smallness of the Fermi velocity $v_F$ Debye screening radius is rather small.
For instance, according to formula (\ref{deb_mass}) for the room temperature and $\epsilon \sim 5$
the screening radius is only $\sim 20 \times$distance between carbon atoms in graphene.

{\it In conclusion}, in this paper we carried out the study of the interaction potential between static charges
in graphene within Monte-Carlo simulation for different dielectric permittivities of substrate $\epsilon$
and various temperatures. To calculate the interaction potential we measured the
correlator of Polyakov lines. At low temperatures the interaction can be satisfactory described by the
Coulomb potential screened by some dielectric permittivity $\epsilon_R$. We determined the dependence
of the $\epsilon_R$ on the dielectric permittivity of substrate. In addition, we determined
the dependence of the renormalized charge squared $\alpha_R$ on the bare one $\alpha_0$
 and showed that at in the semimetal phase the
$\alpha_R$ can be well described by one loop formula.

At larger temperatures
the interaction potential deviates from Coulomb. The main result of this paper is that
for all temperatures and dielectric permittivities the interaction
 can be well desribed by the potential of Debye
screening of two-dimensional plasma of fermion excitations. It is shown that at low temperature Debye mass $m_D$
plays a role of order parameter of the insulator-semimetal phase transition.
At small dielectric permittivity of substrate,  $m_D$  equals zero within
the accuracy of the calculation, what means that the interaction potential
is Coulomb. At $\epsilon \sim 4-5$ Debye mass becomes nonzero,
abruptly reaching the regime of two-dimensional plasma.
The interaction in this region is due to Debye potential.  Thus the study of Debye screening mass
allows to determine the position of the insulator-semimetal phase transition, which
takes place at $\epsilon \sim 4-5$. At large temperatures $m_D$ is not zero for any values of the $\epsilon$.
It is smoothly rising function of $\epsilon$ which is saturated at $\epsilon \sim 4-5$.
In the semimetal phase for all temperatures studied in this paper
  Debye mass can be rather well described by the formula for two-dimensional plasma of fermions
excitations, where the interactions between excitations are accounted by the renormalization of the
charge squared $\alpha_R$ and the Fermi velocity $v_F^R$.

\begin{acknowledgments}
 The authors are grateful to Prof. Mikhail Zubkov for interesting and useful discussions. The work was supported by Grant RFBR-11-02-01227-a and by the Russian Ministry of Science and Education, under contract No. 07.514.12.4028. Numerical calculations were performed at the  ITEP
system Graphyn and Stakan (authors are much obliged to A.V. Barylov, A.A. Golubev, V.A. Kolosov, I.E. Korolko, M.M. Sokolov for the help), the MVS 100K at Moscow Joint Supercomputer Center and at Supercomputing Center of the Moscow State University.
\end{acknowledgments}

\appendix
\section{Debye screening in graphene}
\label{appendix}

This section is devoted to the derivation of the potential of Debye screening in graphene. An important
difference between graphene and usual three dimensional electromagnetic plasma is that free
charges in graphene are two-dimensional. It will be seen below that this property leads to the change of the exponential screening
to power screening.

Suppose that positive charge Q is  located at the origin of coordinates. It is clear that quasiparticles with positive
charge $+e$ repell from the charge Q. The two-dimensional density of positive quasiparticles on graphene plane can be found from Boltzmann
distribution
\beq
n_+(r) = n \exp {\biggl (  - \frac {e \varphi(r)} {T} \biggr )},
\label{bolt1}
\eeq
where $n$ is a density of positive quasiparticles at infinity, $\varphi(r)$ is the potential
which is created by the charge $Q$. Analogously, negative quasiparticles attract to the $Q$ and
their density on graphene plane $n_-(r)$ can be found as follows
\beq
n_-(r) = n \exp {\biggl (   \frac {e \varphi(r)} {T} \biggr )},
\label{bolt2}
\eeq
Evidently, the charge density at distance $r$ is
\beq
\rho(r) = e ( n_+(r) - n_-(r) ) \simeq - 2 n \frac {e^2 \varphi(r)} {T}.
\eeq
In last equation it was assumed that $e \varphi \ll  T$.  Taking into account nonzero charge
density  the $\rho(r)$,  one can write Maxwell equation
\beq
-\Delta \varphi + \frac {8 \pi n e^2} T \delta(z) \varphi = 4 \pi Q \delta^3 ( \vec r).
\eeq
Note that the delta-function $\delta(z)$ in the second term takes into account the fact that the charges
are located on the graphene plane $z=0$. The solution of the Maxwell equation on the graphene plane
can be written as follows
\beq
\varphi(r) & = & Q \int \frac {d^2 p} {(2 \pi)}  \frac  { e^{i \vec p \vec r} } {|p| + m_D} \nonumber \\ &=& \frac Q r \int_0^{\infty} d \xi \frac {e^{- (m_D r) \xi}} { (1+\xi^2)^{3/2} } \xi ,
\label{deb_cont} \\
\frac {m_D} T & = & \frac {4 \pi e^2 n} {T} = \frac {2 \pi^2} 3 \frac {e^2} {v_F^2} ,
\label{deb_mass}
\eeq
here $\vec p = (p_x, p_y)$. It should be noted here that one can use Fermi distribution
instead of Boltzmann distributions (\ref{bolt1}),  (\ref{bolt2})
\beq
n_{\pm} (r) = \int \frac {d^2 p} {(2 \pi)^2} \frac 1 { \exp { \bigl [ (v_F | \vec p| \pm e \varphi(r) )/T \bigr ]} +1},
\eeq
expand them in the ratio $e \varphi (r)/T$ and repeat all the above steps. This leads to the
same expression for the potential $\varphi(r)$ (\ref{deb_cont}) but with different Debye mass
\beq
\frac {m_D} T =  8 \log {2} \frac {e^2} {v_F^2},
\label{deb_mass1}
\eeq
which is by 15 \% smaller than Debye mass in equation (\ref{deb_mass}).

The solution $\varphi(r)$ satisfies the following limits
\begin{equation}
\nonumber
\varphi(r) =
\left \{
\begin{array}{cc}
\frac {Q} r,~~~  ( r m_D ) \ll 1 \\
\frac {Q} r \frac 1 {(m_D r)^2},~~~  ( r m_D ) \gg 1.
\end{array}
\right.
\end{equation}
Thus at large distances Debye screening leads to $\sim 1/r^3$ decrease of the potential.
It causes no difficulties to write lattice version of the potential (\ref{deb_cont}) on the graphene plane
\beq
\label{deb_lat}
V_D( \vec r) = 4 \pi e^2 \sum_{n_1, n_2} \frac {f(p_1,p_2)} {1+ 2 m_D (L_s a_s)^2 f(p_1,p_2)}  e^{i \vec p \vec r}, \\ \nonumber
f(p_1,p_2) = \frac 1 {4 L_s^3 a_s} \sum_{n_3} \frac 1 {\sum_i \sin^2(p_i a_s/2) },~~ p_i=\frac {2 \pi} {L_s a_s} n_i.
\eeq
In formula (\ref{deb_lat}) the integers $n_1$, $n_2$, $n_3$ run the values $0,1,..,L_s-1$, except the
case $n_1=n_2=0$.

\end{document}